\documentclass[pra,twocolumn]{revtex4-2}

\usepackage{graphicx}
\usepackage{mathrsfs}

\usepackage{subfigure}

\usepackage{amsmath}

\usepackage{amstext}
\usepackage{amssymb}
\usepackage{amsbsy}

\usepackage{amsthm}
\usepackage{graphicx}

\usepackage{epstopdf}

\usepackage{color}
\usepackage{diagbox}

\definecolor{Dgreen}{RGB}{0, 100, 0}
\usepackage{url}
\usepackage[colorlinks]{hyperref}
\hypersetup{%
	plainpages=true,
	breaklinks=true, 
	hypertexnames=false, 
	pageanchor=true,
	colorlinks=true,
	linkcolor={blue},
	citecolor={blue},
	urlcolor={blue},
	anchorcolor={black}
}

\makeatother

\begin{document}
\title{Non-Markovian environment induced chaos in optomechanical system}
\author{You-Lin Xiang}
\affiliation{Fujian Key Laboratory of Quantum Information and Quantum Optics (Fuzhou
University), Fuzhou 350108, China}
\affiliation{Department of Physics, Fuzhou University, Fuzhou 350108, China}
\author{Xinyu Zhao}
\thanks{xzhao@fzu.edu.cn}
\affiliation{Fujian Key Laboratory of Quantum Information and Quantum Optics (Fuzhou
University), Fuzhou 350108, China}
\affiliation{Department of Physics, Fuzhou University, Fuzhou 350108, China}
\author{Yan Xia}
\thanks{xia-208@163.com}
\affiliation{Fujian Key Laboratory of Quantum Information and Quantum Optics (Fuzhou
University), Fuzhou 350108, China}
\affiliation{Department of Physics, Fuzhou University, Fuzhou 350108, China}
\begin{abstract}
In traditional research, chaos is frequently accompanied by non-linearity,
which typically stems from non-linear interactions or external driving
forces. However, in this paper, we present the chaotic behavior that
is completely attributed to the non-linear back-reaction of non-Markovian
environment. To be specific, we derive the dynamical equations of an optomechanical system and demonstrate that the non-linearity (cause of chaos) in the equations arises entirely from the time-domain convolutions (TDCs) induced by non-Markovian corrections. Under Markovian conditions, these TDCs are reduced into constants, thereby losing the nonlinearity and ultimately leading to the disappearance of chaos. Furthermore, we also observe chaos generation in the absence of optomechanical couplings, which further confirms that the non-Markovian effect is the sole inducement of chaos and the environmental parameters play
important roles in the generation of chaos. We hope these results
may open a new direction to investigate chaotic dynamics purely caused
by non-Markovian environments. 
\end{abstract}
\maketitle

\section{Introduction}

\label{I} Chaos has been widely studied in many fields, such as neural
networks~\cite{Laje2013NN}, extreme event statistics~\cite{Selmi2016PRL,Coulibaly2017PRA},
and biophysics of chaotic self-organization~\cite{Eskov2017B}. The
investigation of chaos not only advances our understanding of fundamental
physics but also offers significant potential for a range of applications~\cite{Argyris2005N,WOS:000072115200040,Qiu2025NJoP,PhysRevLett.103.024102,Yu2025OE,Xu2025OE,Hirano:10,Alessio2016AiP,Gu2025AQT,Zhu2023FR,Wu2017NC,Fisher2023ARoCMP,Chen2024CP,Bouchez2021PRE,Cao2022PRR,Chen2022PRA,Cornelius2022PRL,Zhao2025,Dowling2023PRL,Li2024PRA,Madiot2021PRA,Martinez2021PRL,Park2023PRB,Polyakov2024PRA,Ray2022PRL,Tikhanovskaya2022PRL,Ustinov2021PRB,Wanzenboeck2021PRA,Xu2021PRR,Zhai2022PRB}.
For example, chaos plays an important role in secret communications~\cite{Argyris2005N,WOS:000072115200040},
rapid generation of high-quality random bit sequences~\cite{PhysRevLett.103.024102,Hirano:10},
and chaos assisted computing~\cite{WOS:000261740400011}.

Traditional theory believes that chaos arises from the nonlinear interactions
within the system~\cite{GoldsteinBook,Lorenz1963JotAS}. Therefore,
a significant amount of research focuses on the physical systems whose
dynamics is governed by nonlinear equations. Optomechanical system
is such an example that exhibits a wealth of nonlinear characteristics~\cite{PhysRevA.83.043826,Yan2025OE,WOS:000273344900032,WOS:000276205000034,WOS:000288170200032,PhysRevA.83.033820,PhysRevLett.105.070403,PhysRevA.79.063801,PhysRevA.102.023707,PhysRevA.104.043521,PhysRevA.106.033509,PhysRevA.106.013526,Zhao:19,PhysRevResearch.4.033102,PhysRevA.100.063827},
originating from the interaction between the cavity mode and the mechanical
oscillator. Therefore, optomechanical systems have become one of the
candidate platforms for the study and exploration of classical and
quantum chaos~\cite{PhysRevA.81.013802,PhysRevA.109.023529,PhysRevA.84.021804,PhysRevLett.114.013601,PhysRevA.101.053851,PhysRevLett.114.253601}.
Over the past few years, extensive theoretical and experimental research
has been conducted on optomechanical systems, for example, Hamiltonian
chaos~\cite{PhysRevA.81.013802}, loss-induced chaos~\cite{PhysRevA.109.023529},
and intermittent chaos~\cite{PhysRevA.101.053851}.

In the studies above and other traditional research~\cite{Zhu2019PRA,PhysRevLett.114.013601,PhysRevA.101.053851,PhysRevLett.114.253601,PhysRevA.84.021804,Ma2023AdP}
on chaos in optomechanical systems, the dynamical equations are always
nonlinear. However, in this paper, we present chaotic behavior in
an optomechanical system purely caused by the back-reaction of the
non-Markovian environment. In our model, the dynamical equation is
in a linear format, the non-linearty only arises from the time-domain
convolutions (TDCs), satisfying a set of non-linear differential equations.
It is proved that those TDCs merely arises from the non-Markovian
corrections to the coefficients in the master equation. Therefore,
the chaotic behavior is an exclusive feature for non-Markovian case.
In Markovian regime, the TDCs are reduced to constants resulting in
the vanish of chaos.

To be specific, we investigate the dynamics of an optomechanical system
with two movable mirrors beyond the Markov approximation. By employing
the non-Markovian quantum state diffusion~(NMQSD) approach~\cite{PhysRevLett.82.1801,PhysRevA.60.91,PhysRevA.58.1699,PhysRevA.69.052115,PhysRevA.69.062107,PhysRevLett.105.240403,PhysRevA.85.040101,PhysRevA.90.052104,Zhao2025},
the non-Markovian master equation is derived. Then, a set of dynamical
equations are derived for physical observables. Although these dynamical
equations are in linear form, the coefficients $F_{i}(t)$ (TDCs)
in the dynamical equations satisfy another set of nonlinear differential
equations when the non-Markovian correction is taken into consideration.

We employ the maximum Lyapunov exponent (LE)~\cite{GoldsteinBook,Benettin1980M,WOLF1985285}
to measure the chaotic behavior in the dynamical process and prove
that chaos is only generated in non-Markovian regime. Analytical results
confirm that both non-linearity and chaotic behavior arise purely
from the TDCs in the dynamical equation, which implies non-Markovian
correction is the only cause of chaos. In another word, the chaotic
behavior is a unique feature in non-Markovian dynamics and never occurs
in Markovian regime.  All these results are also demonstrated in numerical
simulations. Furthermore, we have also excluded the influence of optomechanical couplings on chaos generation. In the absence of optomechanical couplings, numerical results demonstrate that non-Markovian effect is the only cause of chaos.

From the perspective of physical pictures, in traditional
research, non-linearity as a necessary condition for chaos comes from
either the external driving force \cite{Ma2014PRA,Jin2014APL} or
intrinsic nonlinear interactions (reflected as nonlinear terms in
the dynamical equations \cite{Ma2023PRE,Lorenz1963JotAS}). In our
research, the non-Markovian back-reaction from the environment becomes
the provider of nonlinear effects. In summary, non-Markovian effects
cause the non-linearity in TDCs, further introduce chaos in the dynamics
of system, even if the dynamical equation is in a linear format.

Last but not least, we have further studied the influence of several environmental
parameters in the chaos generation process, such as the memory time,
dissipation rate, and central frequency of the environment. Our research
indicates that environmental factors can have a crucial impact on
the generation of chaos. The non-Markovianity of the environment acts
as a switch to control the generation and disappearance of chaos.
We aspire that the findings presented in this paper could pave a novel
way to delve into chaotic dynamics solely stemming from non-Markovian
effects.

The paper is organized as follows.~In~Sec.~\ref{II}, we derive
the non-Markovian master equation for the double-mirror optomechanical
system. In~Sec.~\ref{III}, we introduce the method of quantifying
chaos and demonstrate that chaos can be only generated in non-Markovian
regime. In~Sec.~\ref{sec:IV}, we show the influence of environmental
parameters. Finally, conclusion and outlook are given in~Sec.~\ref{sec:V}.

\section{NON-MARKOVIAN DYNAMICS OF DOUBLE-MIRROR OPTOMECHANICAL SYSTEM}

\label{II}

\subsection{Double-mirror optomechanical system with non-Markovian environment}

\begin{figure} [t]
	\noindent\centering
	\includegraphics[width=1\columnwidth]{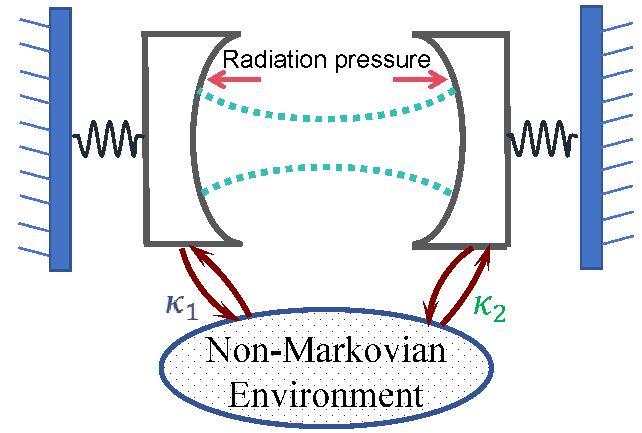} 
	\caption{Schematic diagram of double-mirror optomechanical system. An F-P cavity
		is coupled to two movable mirrors through the radiation pressure.
		Both mirrors are coupled to a non-Markovian common environment.}
	\label{fig1} 
\end{figure}

As shown in~Fig.~\ref{fig1}, the system we considered is a double-mirror
optomechanical system, which is an Fabry-Perot cavity with two reflective
movable mirrors. The motions of two mirrors can be modeled as quantum
harmonic oscillators, while the cavity field can be described by a
light field. Therefore, the total Hamiltonian is written as (assuming
$\hbar=1$) 
\begin{equation}
\hat{H}_{\textrm{tot}}=\hat{H}_{\textrm{S}}+\hat{H}_{\textrm{B}}+\hat{H}_{\textrm{int}},\label{eq:Htot}
\end{equation}
where 
\begin{equation}
\hat{H}_{\textrm{S}}=\omega_{c}\hat{a}^{\dagger}\hat{a}+\sum_{j=1}^{2}\omega_{j}(\hat{p}_{j}^{2}+\hat{q}_{j}^{2})+\sum_{j=1}^{2}G_{j}\hat{a}^{\dagger}\hat{a}\hat{q}_{j},\label{eq:Hs}
\end{equation}
is the Hamiltonian for the double-mirror optomechanical system~\cite{Law1995PRA}.
The position and momentum operators for the two mirrors are indicated
by $\hat{p}_{j}$ and $\hat{q}_{j}$ $(j=1,2)$, respectively, and
the creation (annihilation) operators of the cavity mode are represented
by $\hat{a}^{\dagger}$ $(\hat{a})$. In Eq.~(\ref{eq:Hs}), the
term $G_{j}\hat{a}^{\dagger}\hat{a}\hat{q}_{j}$ represents the interaction
between the radiation field and the two mirrors. Through radiation
pressure, the positions of the two mirrors are coupled to the photon
numbers in the cavity field with the coupling strength $G_{1}$ and
$G_{2}$, respectively.

A general bosonic common environment can be modeled as 
\begin{equation}
\hat{H}_{\textrm{B}}=\sum_{i}\nu_{i}\hat{b}_{i}^{\dagger}\hat{b}_{i},\label{eq:HB}
\end{equation}
where $\hat{b}_{i}$ is the annihilation operator of the $i^{{\rm th}}$
mode. The interaction between system and environment can be described
as 
\begin{equation}
\hat{H}_{\textrm{int}}=\sum_{i}g_{i}(\kappa_{1}\hat{q}_{1}+\kappa_{2}\hat{q}_{2})(\hat{b}_{i}^{\dagger}+\hat{b}_{i}),\label{eq:Hint1}
\end{equation}
where $(\kappa_{1}\hat{q}_{1}+\kappa_{2}\hat{q}_{2})$ describes the
collective interaction between system and environment with the dimensionless
coefficients $\kappa_{1}$, $\kappa_{2}$, respectively. The coupling
strength to the $i^{{\rm th}}$ mode of the environment is $g_{i}$.

Here,we assume that the dissipation of the movable mirrors is the
primary dissipative channel of the system and two mirrors are coupled
to a common environment. This is the most common case because a massive
object (mirror) loss its coherence much faster than light field. In
practical cases, the leakage of the cavity is another source of dissipation.
The corresponding discussion on cavity leakage is shown in Appendix~\ref{Sec:AppLeakCavity}.
Besides, the temperature of the environment also has a significant
impact on the dynamics of the open system. The solution to the finite
temperature case is presented in ~\ref{Sec:AppFiniteT}.

The properties of the environment are mainly determined by the correlation
function (see Appendix~\ref{Sec:AppendixA1} for details) 
\begin{equation}
\alpha(t,s)=\sum_{i}|g_{i}|^{2}e^{-i\nu_{i}(t-s)}.\label{eq:CF}
\end{equation}
Consider a continuous distribution of mode frequency, the summation
over mode ``$i$'' can be replaced by the integration over frequency
``$\nu$'' as 
\begin{equation}
\alpha(t,s)=\int_{0}^{\infty}J(\nu)e^{-i\nu(t-s)}d\nu,
\end{equation}
where $J(\nu)=|g(\nu)|^{2}$ is the spectrum density describing the
coupling strength to the environmental modes with eigen-frequency
$\nu$. A commonly studied case is that the coupling strengths for
different modes are identical, namely $J(\nu)=\Gamma$, the correlation
function becomes a delta-function $\alpha(t,s)=\frac{\Gamma}{2}\delta(t,s)$.
This indicates the environment is memory-less, as it is often called
``Markovian'' environment.

In the most general case, the spectrum density $J(\nu)$ can be arbitrarily
complicated. Here, we explore a widely studied case called the Lorentzian
spectrum density~\cite{PhysRevLett.103.210401,PhysRevB.78.235311,PhysRevA.105.042217,Safavi-Naeini_2013}
\begin{equation}
J(\nu)=\frac{\Gamma\gamma^{2}/2\pi}{(\nu-\Omega)^{2}+\gamma^{2}},\label{eq:SD}
\end{equation}
where $\Omega$ is the central frequency of the environment. The coupling
to the mode with $\nu=\Omega$ is the strongest, while the couplings
are much weaker when the mode frequency $\nu$ is far away from $\Omega$
($|\nu-\Omega|\gg\gamma$). Such a spectrum density $J(\nu)$ corresponds
to the so-called Ornstein-Uhlenbeck (O-U) correlation function (assuming
$\Gamma=1$) 
\begin{equation}
\alpha(t,s)=\int_{0}^{\infty}J(\nu)e^{-i\nu(t-s)}d\nu=\frac{\Gamma\gamma}{2}e^{-(\gamma+i\Omega)|t-s|},\label{eq:CFOU}
\end{equation}
where the residue theorem is used in the calculation of the integral.
Now, the physical meaning of the parameter $\gamma$ is clear. Noticing
that $1/\gamma$ has the unit of time, $\tau=1/\gamma$ can somehow
measure the memory time of the environment. When two points ``$t$''
and ``$s$'' in the time domain are separated much larger than the
memory time $|t-s|\gg\tau$, their correlation is approaching zero,
$\alpha(t,s)\rightarrow0$. In O-U correlation function, the transition
from Markovian to non-Markovian regime can be observed by tuning a
single parameter $\gamma$. When $\gamma\rightarrow\infty$, the environment
exhibits Markovian properties, while $\gamma$ is small, the non-Markovian
corrections become non-negligible.

Typically, other types of correlation functions can be decomposed
into combinations of several O-U correlation functions, since an arbitrary
function can be expanded into an exponential Fourier series. As an
example given in Ref.~\cite{PhysRevA.105.042217}, the widely used
$1/f$ noise can be decomposed into many O-U noises. Here, we chose
the O-U correlation function based on the need to show the transition
from a Markovian to a non-Markovian regime by tunning a single parameter
$\tau=1/\gamma$.

\subsection{Master equation and dynamical equations}

By introducing normalized (Bargmann) coherent state~\cite{BARGMANN1971221,Gardiner2010} $|z_{i}\rangle=e^{z_{i}\hat{b}_{i}^{\dagger} } |0 \rangle$ as the basis to expand the total state vector $|\psi_{\textrm {tot}}(t)\rangle$ (resides in the joint Hilbert space $\mathcal{H}_ {\textrm{ sys}} \otimes \mathcal{H}_{\textrm {env}}$),
one can define a stochastic state vector belonging to the system's Hilbert space $\mathcal{H}_{\textrm{sys}}$. Such a stochastic state vector is formally defined as $|\psi_{t}(z^{*})\rangle \equiv \langle z| \psi_{\textrm {tot}} (t)\rangle$,
where $|z\rangle\equiv\prod_{i}|z_{i}\rangle$ represents the collective
coherent states of all the modes in environments (only residing in the environment's Hilbert space $\mathcal{H}_{\textrm{env}}$). The inner product $\langle z|\psi_{\textrm{tot}}\rangle$ projects the total state onto the environmental modes, thereby stripping away environmental degrees of freedom and retaining only the system’s quantum state. According to the
Schr\"{o}dinger equation for the total system, the equation for the
stochastic state vector $|\psi_{t}(z^{*})\rangle$ can be derived
as (see Appendix~\ref{Sec:AppendixA1} for details) 
\begin{eqnarray}
 &  & \frac{\partial}{\partial t}|\psi_{t}(z^{*})\rangle=[-i\hat{H}_{{\rm S}}+(\kappa_{1}\hat{q}_{1}+\kappa_{2}\hat{q}_{2})z_{t}^{*}\nonumber \\
 &  & -(\kappa_{1}^{*}\hat{q}_{1}+\kappa_{2}^{*}\hat{q}_{2})\int_{0}^{t}ds\alpha(t,s)\frac{\delta}{\delta z_{s}^{*}}]|\psi_{t}(z^{*})\rangle,\label{eq:QSD}
\end{eqnarray}
which is called the NMQSD equation~\cite{PhysRevA.58.1699,PhysRevLett.82.1801}.
In~Eq.~(\ref{eq:QSD}), $z_{t}^{*}\equiv-i\sum_{k}{g_{k}z_{k}^{*}e^{i\nu_{k}t}}$
is a stochastic noise satisfying $M\{z_{t}\}=M\{z_{t}z_{s}\}=0$ and
$M\{z_{t}z_{s}^{*}\}=\alpha(t,s)$ where $M\{\cdot\}\equiv\int\dfrac{d^{2}z}{\pi}e^{-|z|^{2}}\{\cdot\}$
stands for the statistical mean over all the noise variable $z$.

Based on the NMQSD Eq.~(\ref{eq:QSD}), the master equation of the
double-mirror optomechanical system can be derived as (see Appendix~\ref{Sec:AppendixA2}
for details) 
\begin{eqnarray}
\frac{\partial}{\partial t}\hat{\rho} & = & -i[\hat{H}_{{\rm S}},\hat{\rho}]\nonumber \\
 & + & \sum_{i=1}^{5}\sum_{j=1}^{2}\left\{ \kappa_{j}F_{i}(t)(\hat{q}_{j}\hat{\rho}\hat{O}_{i}-\hat{\rho}\hat{O}_{i}\hat{q}_{j})+{\rm H.c.}\right\} ,\label{eq:MEQ}
\end{eqnarray}
where the operator $\hat{O}_{i}$ are 
\begin{equation}
\hat{O}_{1}=\hat{q}_{1},\;\hat{O}_{2}=\hat{q}_{2},\;\hat{O}_{3}=\hat{p}_{1},\;\hat{O}_{4}=\hat{p}_{2},\;\hat{O}_{5}=\hat{a}^{\dagger}\hat{a},
\end{equation}
and the coefficients $F_{i}(t)=\int_{0}^{t}\alpha(t,s)f_{i}(t,s)ds$
$(i=1,2,3,4,5)$ are actually some TDCs ($f_{i}$ are governed by
a set of differential equations given in Appendix \ref{Sec:AppendixA1}).
These TDCs $F_{i}(t)$ satisfy the following equations 
\begin{widetext}
\begin{eqnarray}
\frac{d}{dt}F_{1} & = & \dfrac{\gamma}{2}\kappa_{1}-\gamma F_{1}-i\Omega F_{1}+2\omega_{1}F_{3}-i\kappa_{1}F_{1}F_{3}-i\kappa_{2}F_{1}F_{4},\nonumber \\
\frac{d}{dt}F_{2} & = & \dfrac{\gamma}{2}\kappa_{2}-\gamma F_{2}-i\Omega F_{2}+2\omega_{2}F_{4}-i\kappa_{1}F_{2}F_{3}-i\kappa_{2}F_{2}F_{4},\nonumber \\
\frac{d}{dt}F_{3} & = & -\gamma F_{3}-i\Omega F_{3}-2\omega_{1}F_{1}-i\kappa_{1}F_{3}F_{3}-i\kappa_{2}F_{3}F_{4},\nonumber \\
\frac{d}{dt}F_{4} & = & -\gamma F_{4}-i\Omega F_{4}-2\omega_{2}F_{2}-i\kappa_{1}F_{4}F_{3}-i\kappa_{2}F_{4}F_{4},\nonumber \\
\frac{d}{dt}F_{5} & = & -\gamma F_{5}-i\Omega F_{5}+G_{1}F_{3}+G_{2}F_{4}-i\kappa_{1}F_{5}F_{3}-i\kappa_{2}F_{5}F_{4}.\label{eq:dF}
\end{eqnarray}
\end{widetext}

Given the master equation~(\ref{eq:MEQ}) with the TDCs determined
by the Eqs.~(\ref{eq:dF}), one can evaluate the mean values of the
physical observables such as the position and momentum. For an arbitrary
operator $\hat{A}$, the mean value $\langle\hat{A}\rangle$ satisfies
the relation
\begin{equation}
\frac{d}{dt}\langle\hat{A}\rangle=\frac{d}{dt}{\rm tr}(\hat{A}\hat{\rho})={\rm tr}(\hat{A}\frac{d}{dt}\hat{\rho})\label{eq:dtA}
\end{equation}
where the term $\frac{d}{dt}\hat{\rho}$ can be obtained from the
master equation~(\ref{eq:MEQ}). Substituting Eq.~(\ref{eq:MEQ})
into Eq.~(\ref{eq:dtA}), the dynamical equations governing the mean
values of position and momentum of two movable mirrors can be derived.
Since our focus is solely on the two mirrors, we are only interested
in four mean values $\langle\hat{p}_{1}\rangle,\langle\hat{p}_{2}\rangle,\langle\hat{q}_{1}\rangle$,
and~$\langle\hat{q}_{2}\rangle$. However, when computing these four
mean values, $\langle\hat{a}^{\dag}\hat{a}\rangle$ is involved in
the calculation in order to obtain a set of closed equations. The
details are presented in Appendix~\ref{Sec:AppendixA2}. Therefore,
we define a vector as~$\overrightarrow{V}\equiv\left[\langle\hat{q}_{1}\rangle,\langle\hat{q}_{2}\rangle,\langle\hat{p}_{1}\rangle,\langle\hat{p}_{2}\rangle,\langle\hat{a}^{\dagger}\hat{a}\rangle\right]^{T}$,
it satisfies a matrix equation 
\begin{equation}
\frac{d}{dt}\overrightarrow{V}=\overleftrightarrow{M}\overrightarrow{V},\label{eq:DE}
\end{equation}
where the coefficient matrix $\overleftrightarrow{M}$ can be written
as 
\begin{widetext}
\begin{equation}
\overleftrightarrow{M}=\left[\begin{array}{ccccc}
0 & 0 & 2\omega_{1} & 0 & 0\\
0 & 0 & 0 & 2\omega_{2} & 0\\
-2\omega_{1}+\Im(\kappa_{1}F_{1}) & \Im(\kappa_{1}F_{2}) & \Im(\kappa_{1}F_{3}) & \Im(\kappa_{1}F_{4}) & -G_{1}+\Im(\kappa_{1}F_{5})\\
-2\omega_{2}+\Im(\kappa_{2}F_{1}) & \Im(\kappa_{2}F_{2}) & \Im(\kappa_{2}F_{3}) & \Im(\kappa_{2}F_{4}) & -G_{2}+\Im(\kappa_{2}F_{5})\\
0 & 0 & 0 & 0 & 0
\end{array}\right],\label{eq:M}
\end{equation}
\end{widetext}

where $\Im(\cdot)$ indicates the imaginary part.

It is well known that chaos often occurs in a physical system whose
dynamical equation contains nonlinear terms. The most famous example
is the Lorentz equation \cite{Lorenz1963JotAS}. Here, we derive a
linear dynamical equation~(\ref{eq:DE}). According to the traditional
theory, it is not likely to observe chaos in such a system. However,
we also notice that the coefficients (TDCs) $F_{i}$ ($i=1$ to $5$)
satisfy a set of non-linear differential equations (\ref{eq:dF}).
This may provide the non-linearity for generating chaos in non-Markovian
regime. Next, we will use analytical and numerical methods to show
that the chaotic motions can be generated in this model purely by
the non-Markovian back-reaction of the environment.

\section{NON-MARKOVIAN ENVIRONMENT INDUCED CHAOS}

\label{III} 

Chaos describes the sensitivity to initial conditions in the dynamics.
A quantitative measure of the exponential divergence of the dynamics
is the Lyapunov exponent (LE) \cite{GoldsteinBook}. If two orbits
are separated by the small distance $\epsilon_{0}$ at the time $t=t_{0}$,
then, at a later time $t$ their separation is given by
\begin{equation}
\epsilon(t)\sim\epsilon_{0}e^{\lambda t},
\end{equation}
where the $\lambda$ is the LE~\cite{GoldsteinBook,WOLF1985285},
quantifying the average growth of an infinitesimally small deviation
of a regular orbit arising from a perturbation. In phase space, LE
is a measure of the convergence and divergence rates of neighboring
trajectories. A positive value of LE indicates chaotic dynamics and
divergence from initial conditions. Conversely, a negative LE indicates
trajectories in phase space converge to a common fixed point, suggesting
regularity in system evolution. In a word, when LE exceeds zero, chaos
occurs in the dynamics of the system.

Here we employ the Wolf's method of phase reconstruction~\cite{WOLF1985285}
to calculate the maximum LE.

The wolf's method is simply divided into the following steps \cite{Ma2023PRE}:

(1) Taking an initial point in phase space as $Y(t_{0})$, and let
its nearest neighbor $Y_{0}(t_{0})$, with the distance between the
two points set as $A(t_{0})$.

(2) Starting from time $t_{0}$, we track the time evolution of these
two points until the distance between them exceeds a specified value
$\varepsilon$ at time $t_{1}$: 
\begin{equation}
A^{\prime}(t_{1})=|Y(t_{1})-Y_{0}(t_{1})|>\varepsilon
\end{equation}

(3) At this point, we keep $Y(t_{1})$ and identify a new closest
neighbor $Y_{1}(t_{1})$ nearby, making sure the distance $|Y(t_{1})-Y_{1}(t_{1})|<\varepsilon$
and the angle between $A(t_{1})$ and $A^{\prime}(t_{1})$ is minimized.

(4) The process is repeated until the end of the time series, accumulating
a total of $M$ iterations in the tracking evolution. The maximum
Lyapunov exponent is then given by: 
\begin{equation}
\lambda_{1}=\frac{1}{t_{M}-t_{0}}\sum_{k=1}^{M}\ln\frac{A_{t_{k}}^{\prime}}{A_{t_{k}-1}}
\end{equation}

\begin{figure} [t]
\noindent\centering \includegraphics[width=1\columnwidth]{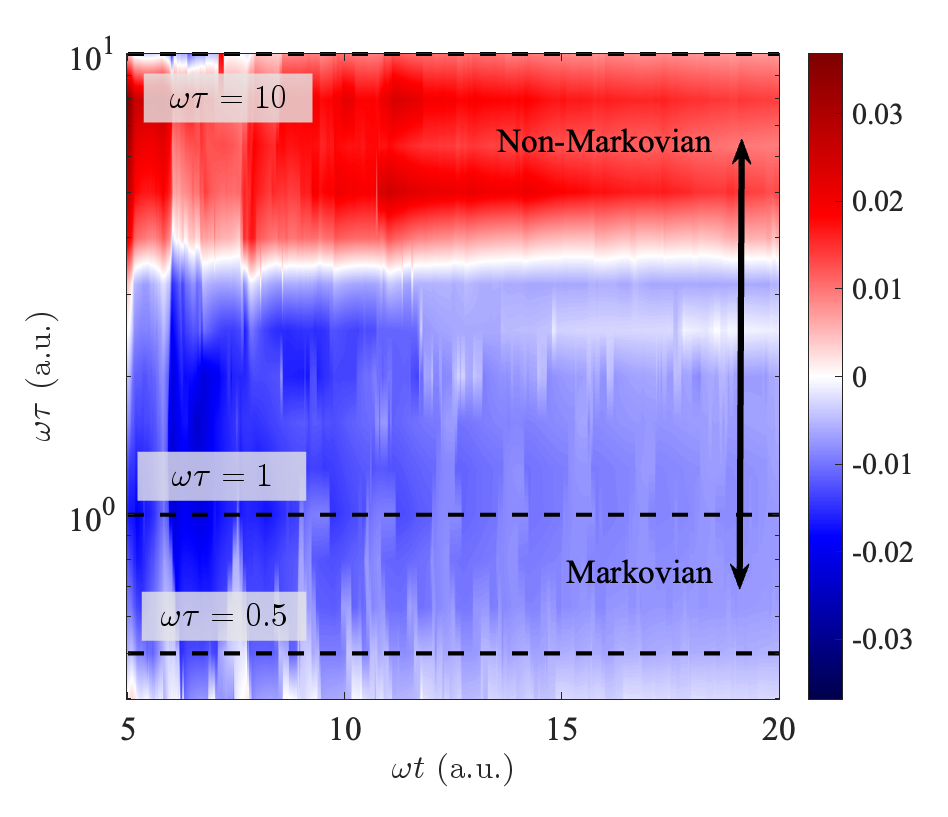}
\caption{The evolution of maximum LE for physical observable
$\langle\hat{q}_{1}\rangle$ with different correlation time $\tau$.
The initial conditions are $\langle\hat{q}_{1}\rangle|_{t=0}=\langle\hat{q}_{2}\rangle|_{t=0}=1.1$
and $\langle\hat{p}_{1}\rangle|_{t=0}=\langle\hat{p}_{2}\rangle|_{t=0}=0$
and $\langle\hat{a}^{\dagger}\hat{a}\rangle|_{t=0}=2$. The parameters
are $\omega_{1}=\omega_{2}=\omega=1$, $\Omega=0$, $G_{1}=G_{2}=1$,
and $\kappa_{1}=\kappa_{2}=1$.}
\label{fig22}
\end{figure}

As we have discussed in Sec.~\ref{II}, the TDCs determine the dynamical
behavior of the system, and the memory time $\tau=1/\gamma$ plays
a crucial role in the time evolution of the TDCs $F_{i}(t)$. To reveal
the impact of $\tau$ on the chaos generation, we plot the maximum
LE of physical observable $\langle\hat{q}_{1}\rangle$ as a function
of $\omega\tau$ and $\omega t$ in Fig.~\ref{fig22}. The ``red''
regions with positive LE indicates a chaotic dynamics, while the ``blue''
regions with negative LE indicates no chaos. From our simulation,
chaos generation strongly depends on the memory time $\tau=1/\gamma$.
Chaos only occurs when $\tau$ is large, i.e., the memory time is
long. The early stage of evolution is not presented because it is
common for a system to take some time to enter a chaotic dynamics.

Taking the fact that the dynamical equation~(\ref{eq:DE}) is in
a linear format, the chaos generated with long memory time $\tau$
in Fig.~\ref{fig22} is somehow counter-intuitive. It is often believed
that chaos is often accompanied by non- linearity. The only nonlinear
factor in the equation is contained in the TDCs that satisfying Eqs.~(\ref{eq:dF}).
Obviously, the TDCs provides the non-linearity for the chaos generation.
Next, we will prove that the TDCs purely originate from the non-Markovian
corrections. In Markovian limit, all the TDCs will be reduced to constants.
As we have discussed in Eq.~(\ref{eq:SD}) and Eq.~(\ref{eq:CFOU}),
small $\tau$ (large $\gamma$) corresponds to a Markovian environment.
Mathematically, it is straightforward to check 
\begin{equation}
\lim_{\gamma\rightarrow\infty}\alpha(t,s)\propto\delta(t,s),
\end{equation}
leading to the TDCs become constants due to the properties of the
$\delta$-function (see Appendix~\ref{Sec:AppendixA1}) 
\begin{equation}
F_{i}(t)\propto\int_{0}^{t}\delta(t,s)f_{i}(t,s)ds=\frac{f_{i}(t,t)}{2}={\rm constant}.
\end{equation}
Therefore, the TDCs do not obey a set of nonlinear differential equations~(\ref{eq:dF})
any more.

In summary, the TDCs provide the non-linearity for chaos generation
in non-Markovian case, resulting in the positive LE region in Fig.~\ref{fig22}.
However, the TDCs are reduced into constants in Markovian limit, resulting
in the disappear of non-linearity, further leading to the disappear
of chaos in the negative LE region in Fig.~\ref{fig22}.

\begin{figure} [t]
	\noindent\centering \includegraphics[width=1\columnwidth]{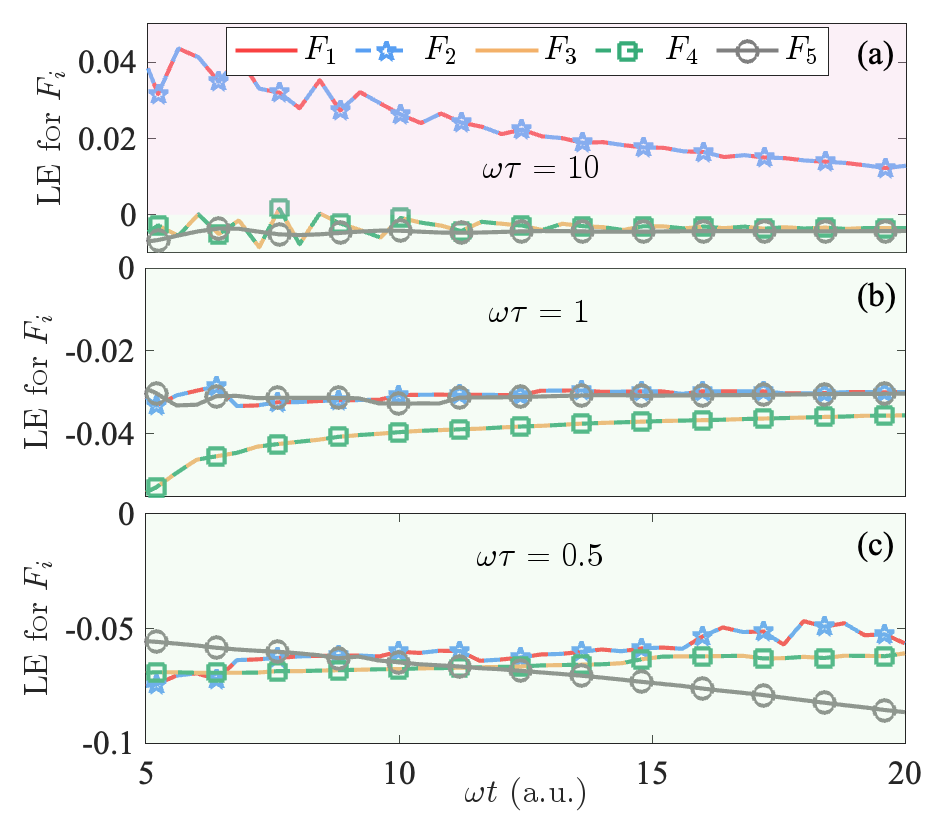}
	\caption{Time evolution of maximum LE of the TDCs $F_{i}$ with different correlation
		time $\tau$. The parameters are the same as Fig.~\ref{fig22}.}
	\label{fig3} 
\end{figure}

To prove the theoretical analysis above, we numerically investigate
the chaotic behaviors of the TDCs $F_{i}$ in Fig.~\ref{fig3}. We
select three particular values of $\omega\tau=0.5$, $\omega\tau=1$,
and $\omega\tau=10$ which are marked by dashed lines in Fig.~\ref{fig22}.
For each given value of $\omega\tau$, The corresponding LE for $F_{i}$
are plotted in Fig.~\ref{fig3}. The chaotic behaviors of $F_{i}$
are coincident with the chaotic behaviors of $\langle q_{1}\rangle$.
To be specific, in the case of long memory time indicated by $\omega\tau=10$,
chaos can be generated in subplot (a). As a comparison, in the cases
of short memory time, presented in subplots (b) and (c), there are
no chaos.

To further illustrate chaos and non-linearity is generated by the
non-Markovian feedback effect other than the optomechanical couplings,
we also numerically simulate the case of $G_{1}=G_{2}=0$. In this
scenario, the possibility that non-linearity is origionated from the optomechanical couplings $G_{1}$ and $G_{2}$ is excluded, since the model is reduced to two harmonic oscillators coupled
to a common environment \cite{Paz2008PRL,Chou2008PRE}.
The figures and detailed discussion are presented in Fig.~\ref{fig6} in Appendix \ref{sec:AppE}, because the numerical plots are similar to Fig.~\ref{fig22} and Fig.~\ref{fig3}. But we would like to emphasize that those results prove the chaotic dynamics we presented is not from optomechanical couplings. This is quite different from most existing studies on chaos in optomechanical systems \cite{Chen2024Entropy,Ma2023AdP,Xiang2025LP}.

From a physical perspective, the non-Markovian back-reaction of the
environment provides non-linearity to the system. Mathematically,
this non-linearity is reflected by the TDCs. Traditional research
focusing on chaos generation in optomechanical system mainly concentrate
on the radiation pressure associated with the parameter $G$, because
this is a non-linear interaction. However, our results reveal an alternative
provider of non-linearity, namely the non-Markovian feedback of the
environment. We hope this opens a path to the study on chaos generation
and we emphasize the important influence of environment should NOT
be ignored.

Certainly, in some limiting cases, non-linearity is not a necessary
condition for chaos. A famous example is the inverted harmonic oscillator
(IHO) \cite{Qu2022a,Hashimoto2020,Qu2022}. However,
in most existing studies, chaos still arises from the properties of
the IHO system itself. Our study shows the possibility of generating
chaos from the interaction between a quantum system and its environment.

\section{INFLUENCE OF ENVIRONMENTAL PARAMETERS}

\label{sec:IV} In Sec.~\ref{III}, we investigate the chaos generation
in double-mirror optomechanical system from a very special angle,
i.e., chaos is purely induced by non-Markovian back-reaction of the
environment. The importance of the environment in chaos generation
has been demonstrated. In this section, we will further reveal the
influence of the environmental parameters.

\subsection{Central frequency}

\label{sua}

\begin{figure}
	\noindent\centering \includegraphics[width=1\columnwidth]{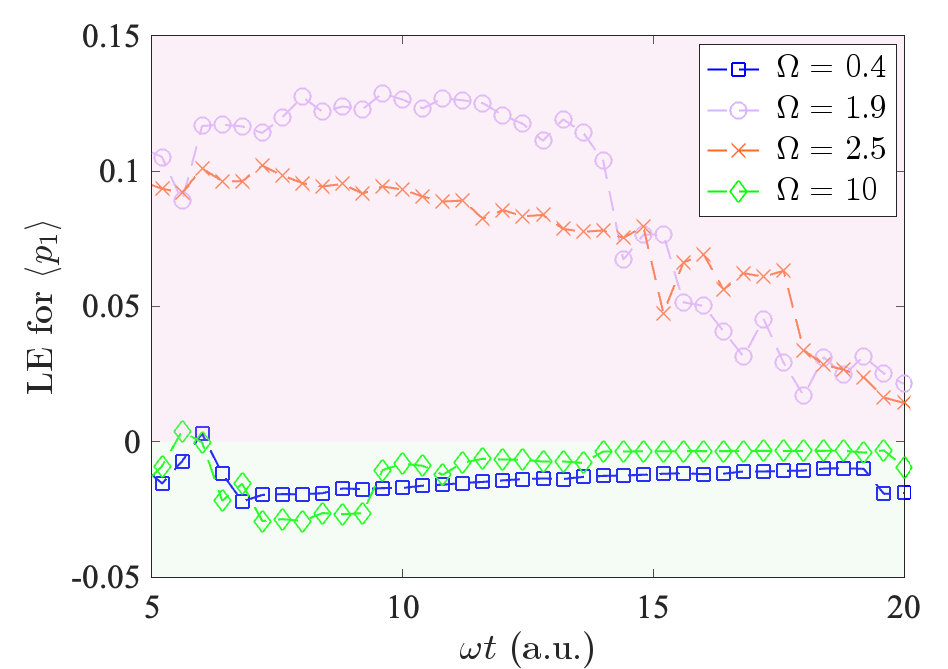}
	\caption{Time evolution of LE of $\langle\hat{p}_{1}\rangle$ with different
		central frequency $\Omega$ of the environment. The memory time $\tau$
		is chosen as $\tau=1$ ($\gamma=1$), and the other parameters are
		the same as Fig.~\ref{fig22}.}
	\label{fig4}
\end{figure}

The properties of the environment are mainly characterized by two
parameters in the correlation function $\alpha(t,s)$, the memory
time $\tau$ and the central frequency $\Omega$. The impact of $\tau$
has been extensively studied in Sec.~\ref{III} where we see the
change of $\tau$ causes the transition from Markovian to non-Markovian
regime. The chaos generation is a unique phenomenon that can be only
observed in non-Markovian regime. Here, we show the influence of another
parameter $\Omega$. In Fig.~\ref{fig4}, the LE of $\langle\hat{p}_{1}\rangle$
is plotted for different values of $\Omega$. It is observed that
only for the case $\Omega\approx2$, LE is positive, indicating chaos
generation. For the other cases, LE is always negative in most regions
of the time evolution.

According to Eq.~(\ref{eq:SD}), $|g(\nu)|^{2}=J(\nu)\propto\frac{\gamma^{2}}{(\nu-\Omega)^{2}+\gamma^{2}}$,
only the bosonic modes with a frequency $\nu\approx\Omega$ have a
strong interaction with the system, the other modes, especially those
far from $\Omega$, interact very weakly with the system. When the
central frequency $\Omega$ is close to a particular frequency, a
strong back-reaction is achieved between the system and environment.
To understand the results in Fig.~\ref{fig4}, one can recall the
resonance phenomenon in classical physics as an analog. When the feedback
has a particular frequency, it has the strongest impact on the system.
Certainly, the numerical results in Fig.~\ref{fig4} reflect a very
complicated multi-mode interactions satisfying a distribution in Eq.~(\ref{eq:SD}),
the analog above is not rigorous. 

If we consider a special case that the environment contains only a
single mode (not the case in Fig.~\ref{fig4}), it would be easier
to capture the physical essence. Mathematically, in this limiting
case, the Hamiltonian in Eq.~(\ref{eq:HB}) is reduced to $\hat{H}_{{\rm B}}=\nu_{1}\hat{b}_{1}^{\dagger}\hat{b}_{1}$
with $\nu_{i}=0$ for all $i\neq1$. Because there is only one mode,
the central frequency is just $\Omega=\nu_{1}$. Besides, the interaction
in Eq.~(\ref{eq:Hint1}) is also reduced to $\hat{H}_{{\rm int}}=g_{1}(\kappa_{1}\hat{q}_{1}+\kappa_{2}\hat{q}_{2})(\hat{b}_{1}^{\dagger}+\hat{b}_{1})$
with $g_{i}=0$ for all $i\neq1$. Now, the dynamics of the two mirrors
is similar to the ``vibration of coupled harmonic oscillator'' which
is widely discussed in the textbooks of classical mechanics~\cite{GoldsteinBook}.
When the frequency of the environment matches the frequency of the
``normal mode'' $\omega_{{\rm normal}}$ of two mirrors, there is
a strong energy exchange between the environment and the two mirrors.
Otherwise, the energy exchange becomes smaller.

If the detuning between $\Omega$ and $\omega_{{\rm normal}}$ is
huge, the influence from the environment is very limited, as if the
two mirrors are not interacting with the environment at all. Therefore,
we only observe the chaos when $\Omega\approx2$ in Fig.~\ref{fig4},
because the two mirrors are approximately decoupled with the environment
in huge detuning cases.

In summary, the central frequency of the environment is also crucial
for the generation of chaos. When the central frequency is far detuned
from the system frequencies, non-Markovian effect is weaken and chaos
can not be generated.

\subsection{Dissipation rate}

\label{sub} 
\begin{figure}[t]
\centering \includegraphics[width=1\columnwidth]{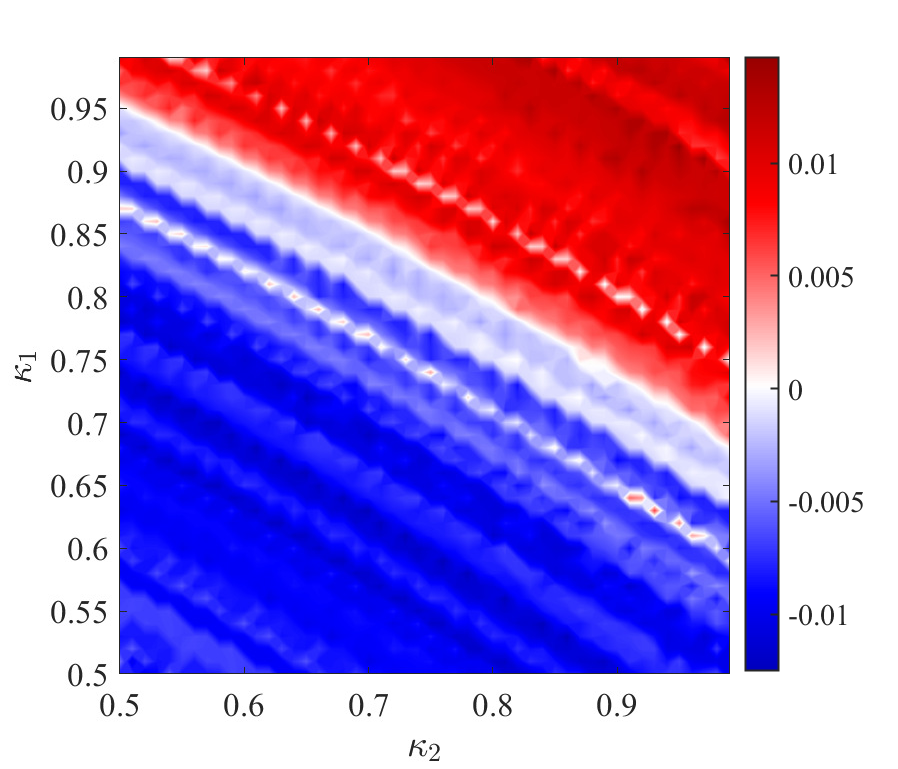} \caption{The average of LE of $\langle\hat{p}_{1}\rangle$ over $\omega t\in[5,20]$.
The parameters are $\gamma=0.5$, $\Omega=0$, $G_{1}=G_{2}=1$, and
\mbox{%
$\omega_{1}=\omega_{2}=1$%
}. The initial conditions are $\langle\hat{q}_{1}\rangle|_{t=0}=\langle\hat{q}_{2}\rangle|_{t=0}=1$,
$\langle\hat{p}_{1}\rangle|_{t=0}=\langle\hat{p}_{2}\rangle|_{t=0}=2$
and $\langle\hat{a}^{\dagger}\hat{a}\rangle|_{t=0}=1$.}
\label{fig5} 
\end{figure}

The dissipation rate is another simple but important
environmental parameter. In Fig.~\ref{fig5}, We investigate the
effects of asymmetric dissipation rates of two mirrors on the chaotic
dynamics. In order to clearly show the optimal parameter range for
chaotic phenomena, we plotted a contour map about the left mirror
dissipation rate ($\kappa_{1}$) and the right mirror dissipation
rate ($\kappa_{2}$). These plots offer a rich dynamics induced by
the parameters $\kappa_{1}$ and $\kappa_{2}$. According to Fig.~\ref{fig5},
chaos generation only occurs when $\kappa_{1}$ and $\kappa_{2}$
are large. As we have analyzed in Sec.~\ref{III}, the nonlinearity
for chaos generation is provided by the back-reaction of the environment,
thus strong couplings $\kappa_{1}$ and $\kappa_{2}$ are necessary.
It is straightforward to understand the absence of chaos in the region
with small $\kappa_{1}$ and $\kappa_{2}$ by considering the limiting
case that $\kappa_{1}=\kappa_{2}=0$. In this case, the environment
is decoupled. The matrix $\overleftrightarrow{M}$ in Eq.~(\ref{eq:M})
will be independent of TDCs $F_{i}(t)$, and the dynamical equation
Eq.~(\ref{eq:DE}) is reduced to a linear equation. Consequently,
the nonlinearity required for chaos disappear and no chaos generation
observed.

The numerical results in Fig.~\ref{fig5} also reflect that the chaos
generation is more sensitive to the parameter $\kappa_{1}$. This
is because we are plotting LE of $\langle\hat{p}_{1}\rangle$. The
dynamics of the left mirror is directly related to $\kappa_{1}$,
and the influence of $\kappa_{2}$ is indirectly. The LE of $\langle\hat{p}_{2}\rangle$
is more sensitive to $\kappa_{2}$ according to our simulation (not
presented).

In summary, we have discussed the influence of the environmental parameters,
including the memory time $\tau$, the central frequency $\Omega$,
and the dissipation rate $\kappa_{i}$. As we have mentioned in Sec.~\ref{I},
most existing studies focus on the nonlinear dynamical equations while
our study focus on a linear dynamical equation. The nonlinearity is
provided by the TDCs which are the consequence of non-Markovian corrections.
Here, we would like to emphasize another difference between our research
and most existing studies, i.e., we focus on the environmental parameters
instead of system parameters. The results presented in this section
show that the environmental parameters are equally important as the
system parameters that are widely discussed in existing studies.

\section{Conclusion}

\label{sec:V} In this paper, we provide an insight into the effect
of non-Markovian environment on chaotic dynamics in optomechanical
systems. We derive the dynamical equation for the double-mirror optomechanical
system. The non-linearity for chaotic dynamics is completely provided
by the TDCs in the equation, which is the consequence of non-Markovian
feedback effect.
The full dynamics in Eq.~(\ref{eq:DE}) combined with Eq.~(\ref{eq:dF}) is, in fact,
non-linear. The linear format of Eq.~(\ref{eq:DE}) is a formal mathematical result that serves to explicitly highlight the origin of nonlinearity: it is not embedded in the structure of the mean-value equation itself, but rather hidden in the TDCs, which are determined by the non-Markovian back-reaction of the environment.
Thus, the chaos observed in this paper
is purely generated by the non-Markovian effect. Both of our analytical
and numerical results prove this conclusion and show that the chaos
disappear in Markovian regime. Then, we further investigate the influence
of the properties of the environment. We show that the environmental
parameters including the memory time, the central frequency and the
dissipation rate are crucial for the chaos generation.

The example presented in this paper open a new window for the investigation
of chaotic behaviors. We have demonstrated that the role of the environment
in chaotic dynamics can be as significant as, or even more important
than, the intrinsic properties of the system itself. Sometimes, the
chaos can be purely generated by the non-Markovian effect of the environment.
Future research may even focus on some linear system in the traditional
theory, but coupled to a non-Markovian environment. The strong back-reaction
from the environment may provide the non-linearity for the chaotic
behaviors. We expect the results presented in this paper will stimulate
greater interest in chaotic phenomena purely induced by non-Markovian
environments.
\begin{acknowledgments}
We thank the fruitful discussion with Dr.~R.~Luo. 
Y. Xia. was supported by the National Natural Science Foundation of China under Grant No. 62471143, the Key Program of National Natural Science Foundation of Fujian Province under Grant No. 2024J02008, and the project from Fuzhou University under Grant No. JG2020001-2. X. Zhao. was supported by Natural Science Foundation of Fujian Province under Grant No. 2022J01548.
\end{acknowledgments}

\appendix

\section{Derivation of master equation}

\subsection{Equations for TDCs}

\label{Sec:AppendixA1}

In this subsection, we show the details of the derivation of the equations
for the TDCs in Eq.~(\ref{eq:dF}) in the main text. First, we start
from the derivation of the NMQSD equation. For the total state vector
$|\psi_{{\rm tot}}\rangle$ describing the quantum state involving
both the system and the environment, it lives in the Hilbert space
of $H_{{\rm tot}}$ and satisfies the Schr\"{o}dinger equation 
\begin{equation}
\frac{d}{dt}|\psi_{{\rm tot}}\rangle=-i\hat{H}_{{\rm tot}}|\psi_{{\rm tot}}\rangle.\label{eq:SEQ}
\end{equation}
Using the Bargemann state basis $|z\rangle$ presented in the main
text, one can take the time-derivative on both side of $|\psi_{t}(z^{*})\rangle\equiv\langle z|\psi_{{\rm tot}}(t)\rangle$.
Noticing the relation $\hat{b}_{i}|z\rangle=z_{i}|z\rangle$ and $\hat{b}_{i}^{\dagger}|z\rangle=\frac{\partial}{\partial z_{i}^{*}}|z\rangle$,
one can obtain Eq.~(\ref{eq:QSD}) in the main text. To solve this
equation, the functional derivative $\frac{\delta}{\delta z_{s}^{*}}$
can be replaced by a time-dependent operator as~$\frac{\delta}{\delta z_{s}^{*}}|\psi_{t}(z^{*})\rangle\equiv\hat{O}(t,s,z^{*})|\psi_{t}(z^{*})\rangle$~with
the initial condition $\hat{O}(t=s,z^{*})=(\kappa_{1}\hat{q}_{1}+\kappa_{2}\hat{q}_{2})$
\cite{PhysRevA.60.91}. Then, the NMQSD equation can be rewritten
as 
\begin{align}
\frac{\partial}{\partial t}|\psi_{t}(z^{*})\rangle & =[-i\hat{H}_{{\rm S}}+(\kappa_{1}\hat{q}_{1}+\kappa_{2}\hat{q}_{2})z_{t}^{*}\nonumber \\
 & -(\kappa_{1}\hat{q}_{1}+\kappa_{2}\hat{q}_{2})\bar{O}(t,z^{*})]|\psi_{t}(z^{*})\rangle,\label{eq:QSDO}
\end{align}
where $\bar{O}(t,z^{*})=\int_{0}^{t}ds\alpha(t,s)\hat{O}(t,s,z^{*})$.
This is the fundamental equation governing the dynamics of the open
system without any approximation. The key to solve Eq.~(\ref{eq:QSDO})
is finding the $\hat{O}$ operator. There is a systematic approach
to find the $\hat{O}$ operator for arbitrary $H_{{\rm S}}$ in Refs.~\cite{PhysRevA.60.91,PhysRevA.86.032116,Zhao2025,PhysRevA.85.042106}.
Here, we only briefly review the procedure of solving $\hat{O}$ operator.

According to the consistency condition $\frac{d}{dt}\frac{\delta}{\delta z_{s}^{*}}|\psi_{t}(z^{*})=\frac{\delta}{\delta z_{s}^{*}}\frac{d}{dt}|\psi_{t}(z^{*})\rangle$,
the operator $\hat{O}$ should satisfy the equation 
\begin{align}
\frac{\partial}{\partial t}\hat{O} & =[-i\hat{H}_{\textrm{S}}+(\kappa_{1}\hat{q}_{1}+\kappa_{2}\hat{q}_{2})z_{t}^{*}\nonumber \\
 & -(\kappa_{1}\hat{q}_{1}+\kappa_{2}\hat{q}_{2})\bar{O},\hat{O}]-\hat{L}^{\dagger}\frac{\delta}{\delta z_{s}^{*}}\bar{O}.\label{eq:CC}
\end{align}
Solving Eq.(\ref{eq:CC}), the exact $\hat{O}$ operator for this
particular model can be determined as 
\begin{equation}
\hat{O}(t,s,z^{\ast})=\sum_{i=1}^{5}f_{i}(t,s)\hat{O}_{i}+\int_{0}^{t}ds^{\prime}f_{6}(t,s,s^{\prime})z_{s^{\prime}}^{\ast},\label{eq:O}
\end{equation}
where the basis operators are 
\begin{equation}
\hat{O}_{1}=\hat{q}_{1},\;\hat{O}_{2}=\hat{q}_{2},\;\hat{O}_{3}=\hat{p}_{1},\;\hat{O}_{4}=\hat{p}_{2},\;\hat{O}_{5}=\hat{a}^{\dagger}\hat{a},\label{eq:Obasis}
\end{equation}
and the coefficients satisfy the following equations 
\begin{align}
\frac{\partial}{\partial t}f_{1} & =2\omega_{1}f_{3}-2i\kappa_{1}F_{1}f_{3}-i\kappa_{1}F_{2}f_{4}+i\kappa_{1}F_{3}f_{1}\nonumber \\
 & +i\kappa_{1}F_{4}f_{2}-i\kappa_{2}F_{1}f_{4}-\kappa_{1}F_{6},\label{eq:f1}
\end{align}
\begin{align}
\frac{\partial}{\partial t}f_{2} & =2\omega_{2}f_{4}-i\kappa_{1}F_{2}f_{3}-i\kappa_{2}F_{1}f_{3}-2i\kappa_{2}F_{1}f_{4}\nonumber \\
 & +i\kappa_{2}F_{3}f_{1}+i\kappa_{2}F_{4}f_{2}-\kappa_{2}F_{6},
\end{align}
\begin{equation}
\frac{\partial}{\partial t}f_{3}=-2\omega_{1}f_{1}-i\kappa_{1}F_{3}f_{3}-i\kappa_{2}F_{3}f_{4},
\end{equation}
\begin{equation}
\frac{\partial}{\partial t}f_{4}=-2\omega_{2}f_{2}-i\kappa_{1}F_{4}f_{3}-i\kappa_{2}F_{4}f_{4},
\end{equation}
\begin{equation}
\frac{\partial}{\partial t}f_{5}=G_{1}f_{3}+G_{2}f_{4}-i\kappa_{1}F_{5}f_{3}-i\kappa_{2}F_{5}f_{4},
\end{equation}
\begin{equation}
\frac{\partial}{\partial t}f_{6}(t,s,s^{\prime})=-i\kappa_{1}f_{3}(t,s)F_{6}(t,s^{\prime})-i\kappa_{2}f_{4}(t,s)F_{6}(t,s^{\prime}),\label{eq:f6}
\end{equation}
where $F_{i}(t)=\int_{0}^{t}\alpha(t,s)f_{i}(t,s)ds\;(i=1,2,3,4,5)$,
and $F_{6}(t,s^{\prime})=\int_{0}^{t}\alpha(t,s)f_{6}(t,s,s^{\prime})ds$.
The boundary conditions for the coefficients are 
\begin{equation}
f_{1}(t,t)=\kappa_{1},\label{f1ini}
\end{equation}
\begin{equation}
f_{2}(t,t)=\kappa_{2},
\end{equation}
\begin{equation}
f_{3}(t,t)=f_{4}(t,t)=f_{5}(t,t)=0,\label{f5ini}
\end{equation}
\begin{equation}
f_{6}(t,t,s^{\prime})=0,\;f_{6}(t,s,t)=i\kappa_{1}f_{3}(t,s)+i\kappa_{2}f_{4}(t,s).
\end{equation}
The differential equations for the coefficients $F_{i}(t)$ can be
computed as 
\begin{eqnarray}
\frac{d}{dt}F_{i}(t) & = & \frac{d}{dt}\int_{0}^{t}\alpha(t,s)f_{i}(t,s)ds\nonumber \\
 & = & \frac{d}{dt}\int_{0}^{t}\frac{\Gamma\gamma}{2}e^{-(\gamma+i\Omega)|t-s|}f_{i}(t,s)ds\nonumber \\
 & = & \frac{\Gamma\gamma}{2}f(t,t)+\int_{0}^{t}\frac{\Gamma\gamma}{2}[\frac{d}{dt}e^{-(\gamma+i\Omega)|t-s|}]f_{i}(t,s)ds\nonumber \\
 &  & +\int_{0}^{t}\frac{\Gamma\gamma}{2}e^{-(\gamma+i\Omega)|t-s|}\frac{d}{dt}f_{i}(t,s)ds.\label{eq:dF-1}
\end{eqnarray}
Substituting Eq.~(\ref{eq:f1})-(\ref{eq:f6}) into Eq.~(\ref{eq:dF-1}),
the differential equations for the TDCs $F_{i}$ in the main text
can be derived as shown in Eq.~(\ref{eq:dF}).

It is worth to note that the dynamical equation is Eq.~(\ref{eq:QSDO})
or the master equation (\ref{eq:MEQ}). The equations (\ref{eq:f1})
to (\ref{eq:f6}) are just for the TDCs. In the main text we only
consider the noise free $\hat{O}$ operator $\hat{O}^{(0)}=\sum_{i=1}^{5}f_{i}(t,s)\hat{O}_{i}$
as an approximate solution. This approximation is irrelevant to the
Hilbert space of the system. Moreover, the accuracy of this approximation
is validated in many existing examples~\cite{PhysRevA.84.032101,PhysRevA.85.042106,PhysRevA.86.032116,PhysRevA.94.012334,Zhao:19}.
Particularly, in Ref. \cite{Xu2014JoPAMaT}, the noise-dependent terms
are proved to have an impact up to the fourth order (or higher) of
$g_{i}$. When $g_{i}$ is not extremely strong, it is always safe
to drop the noise-dependent terms.

\subsection{The non-Markovian master equation}

\label{Sec:AppendixA2} In this subsection, we show the detailed derivation
of the master equation (\ref{eq:MEQ}) in the main text. The density
operator $\rho$ can be constructed by taking the statistical average
over all the possible realization of the stochastic state vector $|\psi_{t}(z^{*})\rangle$
as 
\begin{equation}
\hat{\rho}_{s}=M\{|\psi_{t}(z^{*})\rangle\langle\psi_{t}(z)|\}=\int\frac{d^{2}z}{\pi}e^{-|z|^{2}}|\psi_{t}(z^{*})\rangle\langle\psi_{t}(z)|.\label{rho}
\end{equation}
Taking the time-derivative on both sides of Eq.~(\ref{rho}), one
obtains 
\begin{align}
\frac{d}{dt}\hat{\rho} & =\frac{d}{dt}M\left\{ |\psi_{t}(z^{*})\rangle\langle\psi_{t}(z)|\right\} \nonumber \\
= & M\left\{ \left[\frac{d}{dt}|\psi_{t}(z^{*})\rangle\right]\langle\psi_{t}(z)|\right\} +M\left\{ |\psi_{t}(z^{*})\rangle\frac{d}{dt}\langle\psi_{t}(z)|\right\} \nonumber \\
= & M\left\{ \left[-i\hat{H}_{{\rm S}}+\hat{Q}z_{t}^{*}-\hat{Q}\bar{O}\right]|\psi_{t}(z^{*})\rangle\langle\psi_{t}(z)|\right\} \nonumber \\
 & +M\left\{ |\psi_{t}(z^{*})\rangle\langle\psi_{t}(z)|\left[iH_{{\rm S}}+\hat{Q}z_{t}-\bar{O}^{\dagger}\hat{Q}\right]\right\} ,
\end{align}
where the notation $\hat{Q}=\kappa_{1}\hat{q}_{1}+\kappa_{2}\hat{q}_{2}$
is used as a collective coupling operator for two mirrors. Since the
operator $\hat{H}_{{\rm S}}$ is independent of the noise $z^{*}$
it is straightforward to obtain 
\begin{align}
 & M\{\hat{H}_{{\rm S}}|\psi_{t}(z^{*})\rangle\langle\psi_{t}(z)|\}\nonumber \\
 & =\hat{H}_{{\rm S}}M\{|\psi_{t}(z^{*})\rangle\langle\psi_{t}(z)|\}=\hat{H}_{{\rm S}}\hat{\rho}.
\end{align}
If the operator $\bar{O}$ is also noise independent (neglecting the
$\hat{O}_{6}$ term), similar results can be obtained. Then, the only
task is to compute the term $M\{z_{t}^{*}|\psi_{t}(z^{*})\rangle\langle\psi_{t}(z)|\}$.
Using the Novikov theorem~\cite{PhysRevA.58.1699,PhysRevA.69.052115,PhysRevA.86.032116},
we have the following results 
\begin{align}
M\{|\psi_{t}(z^{*})\rangle\langle\psi_{t}(z)|z_{t}\} & =M\{\bar{O}|\psi_{t}(z^{*})\rangle\langle\psi_{t}(z)|\},\nonumber \\
M\{z_{t}^{*}|\psi_{t}(z^{*})\rangle\langle\psi_{t}(z)|\} & =M\{|\psi_{t}(z^{*})\rangle\langle\psi_{t}(z)|\bar{O}^{\dagger}\}.
\end{align}
Finally, the master equation can be derived as shown in Eq.~(\ref{eq:MEQ})
in the main text.

In the derivation of the master equation, the $z_{t}^{*}$-noise dependent
term associated with $f_{6}$ can be neglected. According to Ref.~\cite{Xu2014JoPAMaT},
the noise dependent term is typically several orders smaller than
other terms. One can also follow the approach presented in Ref.~\cite{PhysRevA.90.052104}
to derive the master equation with noise dependent term.

\section{The differential equations for physical observables}

\label{sec:AppMean} In odder to compute LE, one has to calculate
the mean values of set of operators. Since $\frac{d}{dt}\langle\hat{A}\rangle={\rm tr}(\hat{A}\frac{d}{dt}\hat{\rho})$,
one can obtain the following equations for the mean values 
\begin{equation}
\frac{d}{dt}\langle\hat{q}_{1}\rangle=2\omega_{1}\langle\hat{p}_{1}\rangle,\label{eq:dq1}
\end{equation}
\begin{equation}
\frac{d}{dt}\langle\hat{q}_{2}\rangle=2\omega_{2}\langle\hat{p}_{2}\rangle,
\end{equation}
\begin{align}
\frac{d}{dt}\langle\hat{p}_{1}\rangle & =-2\omega_{1}\langle\hat{q}_{1}\rangle-G_{1}\langle\hat{a}^{\dagger}\hat{a}\rangle\nonumber \\
 & +i\sum_{i=1}^{5}(-\kappa_{1}F_{i}\langle\hat{O}_{i}\rangle+\kappa_{1}^{*}F_{i}^{*}\langle\hat{O}^{\dagger}\rangle),\label{eq:dp1}
\end{align}
\begin{align}
\frac{d}{dt}\langle\hat{p}_{2}\rangle & =-2\omega_{1}\langle\hat{q}_{2}\rangle-G_{2}\langle\hat{a}^{\dagger}\hat{a}\rangle\nonumber \\
 & +i\sum_{i=1}^{5}(-\kappa_{2}F_{i}\langle\hat{O}_{i}\rangle+\kappa_{2}^{*}F_{i}^{*}\langle\hat{O}^{\dagger}\rangle),\label{eq:dp2}
\end{align}
\begin{equation}
\frac{d}{dt}\langle\hat{a}^{\dagger}\hat{a}\rangle=0.\label{eq:dada}
\end{equation}
To investigate the dynamics of the two mirrors the first four equations
are enough. However, due to the optomechnical interaction, the photon
number $\langle\hat{a}^{\dagger}\hat{a}\rangle$ is involved in Eq.~(\ref{eq:dp1})
and Eq.~(\ref{eq:dp2}). In order to solve $\langle\hat{p}_{1}\rangle$
and $\langle\hat{p}_{2}\rangle$, one has to also include $\langle\hat{a}^{\dagger}\hat{a}\rangle$
to form a set of closed equations which can be rewritten in a matrix
form as shown in Eq.~(\ref{eq:DE}).

\section{NMQSD equation in the presence of cavity leakage}

\label{Sec:AppLeakCavity} In the main text, we assume the main decoherence
channel is the damping of the two mirrors. Although the leakage of
the cavity is relatively weak in most cases, it is still a possible
source of decoherence. In this section, we present the procedure to
derive the dynamical equations in the case that the cavity leakage
is dominant.

When the cavity leakage is the primary system-environment interaction,
the interaction Hamiltonian can be written as 
\begin{equation}
\hat{H}_{\text{int}}=\sum_{i}g_{i}\hat{a}\hat{b}_{i}^{\dagger}+\text{H.c.}.
\end{equation}
Similarly, one can also introduce the same stochastic state vector
$|\psi_{t}(z^{*})\rangle=\langle z^{*}|\psi_{\text{tot}}(t)\rangle$
with stochastic variables $z^{*}$. Then, the NMQSD equation for this
system can be derived as
\begin{equation}
\frac{\partial}{\partial t}|\psi_{t}(z^{*})\rangle=\left[-i\hat{H}_{\text{S}}+\hat{a}z_{t}^{*}-\hat{a}^{\dagger}\int_{t_{0}}^{t}ds\alpha(t,s)\frac{\delta}{\delta z_{s}^{*}}\right]|\psi_{t}(z^{*})\rangle,\label{eq:QSDLeak}
\end{equation}
with 
\begin{equation}
z_{t}^{*}=-i\sum_{i}g_{i}z_{m}^{*}e^{i\omega_{i}t},
\end{equation}
The correlation function is 
\begin{equation}
\alpha(t,s)=\sum_{i}|g_{i}|^{2}e^{-i\omega_{i}(t-s)}.
\end{equation}
To solve this equation, the functional derivative $\frac{\delta}{\delta z_{s}^{*}}$
can be replaced by a time-dependent operator as $\frac{\delta}{\delta z_{s}^{*}}|\psi_{t}(z^{*})\rangle\equiv\hat{O}(t,s,z^{*})|\psi_{t}(z^{*})\rangle$
with the initial condition $\hat{O}(t=s,z^{*})=\hat{a}$~\cite{PhysRevA.60.91}.
Then, the NMQSD equation can be rewritten as 
\begin{align}
\frac{\partial}{\partial t}|\psi_{t}(z^{*})\rangle & =[-i\hat{H}_{{\rm S}}+\hat{a}z_{t}^{*}-\hat{a}^{\dagger}\bar{O}(t,z^{*})]|\psi_{t}(z^{*})\rangle,
\end{align}
where $\bar{O}(t,z^{*})=\int_{0}^{t}ds\alpha(t,s)O(t,s,z^{*})$. According
to the consistency condition $\frac{d}{dt}\frac{\delta}{\delta z_{s}^{*}}|\psi_{t}(z^{*})\rangle = \frac{\delta}{\delta z_{s}^{*}}\frac{d}{dt}|\psi_{t}(z^{*})\rangle$,
the operator $\hat{O}$should satisfy the equation 
\begin{align}
\frac{\partial}{\partial t}\hat{O} & =[-i\hat{H}_{\textrm{S}}+\hat{a}z_{t}^{*}-\hat{a}^{\dagger}\bar{O},\hat{O}]-\hat{a}^{\dagger}\frac{\delta}{\delta z_{s}^{*}}\bar{O}.\label{aeq}
\end{align}
Solving Eq. (\ref{aeq}), the $\hat{O}$ operator for this particular
model can be determined as 
\begin{equation}
\hat{O}(t,s,z^{\ast})=\sum_{i=1}^{5}x_{i}(t,s)\hat{O}_{i},\label{aeq:O}
\end{equation}
where the basis operators are 
\begin{equation}
\hat{O}_{1}=\hat{a},\;\hat{O}_{2}=\hat{q}_{1}\hat{a},\;\hat{O}_{3}=\hat{p}_{1}\hat{a},\;\hat{O}_{4}=\hat{q}_{2}\hat{a},\;\hat{O}_{5}=\hat{p}_{2}\hat{a},\label{aeq:Obasis}
\end{equation}
Substituting Eq.~(\ref{aeq:O}) into Eq.~(\ref{aeq}), one can obtain
set of equation for the coefficients $x_{i}(t,s)$ which is similar
to the differential equations shown in Eq.~(\ref{eq:f1})-(\ref{eq:f6}).
Then, one can obtain the TDCs by following the similar procedure shown
in Appendix~\ref{Sec:AppendixA1}.

Finally, the master equation can be derived as 
\begin{equation}
\frac{d}{dt}\hat{\rho}=-i[\hat{H}_{\text{S}},\hat{\rho}_{}]+[\hat{a},\hat{\rho}_{}\bar{O}_{}^{\dagger}]-[\hat{a}^{\dagger},\bar{O}_{}\hat{\rho}_{t}].
\end{equation}

\section{NMQSD equation for finite temperature}

\label{Sec:AppFiniteT} The temperature is also an important factor
in the dynamics of the open system. In the main text, we assume the
system is initially coupled to a vacuum state of the environment,
which implies the initial temperature is zero. This is because the
central topic of this paper is the non-Markovian feedback effect.
The modification on the feedback caused by finite temperature can
be studied in a future research. Here, in this section, we briefly
show the procedure of deriving the dynamical equations in the case
of finite temperature.

The general routine of solving finite temperature case is already
set up in~\cite{PhysRevA.69.062107}. The fundamental idea is to
introduce another fictitious bath $H_{C}=-\sum_{k}\omega_{k}c_{k}^{\dagger}c_{k}$,
separated from all the other systems and environments. Thus, the fictitious
bath does not affect the evolution of the original Hamiltonian. Then,
by introducing Bogoliubov transformations, the initial thermal state
for real bath $\rho_{B}(0)=e^{-\beta H_{B}}/Z$ is transformed into
an effective vacuum state in the new basis. Eventually, a finite temperature
problem with one real bath is mapped into a zero-temperature problem
with two effective baths. According to~\cite{PhysRevA.69.062107},
the NMQSD equation for finite temperature case is 
\begin{align}
\frac{\partial}{\partial t}|\psi_{t}(z^{*},w^{*})\rangle & =[-i\hat{H}_{S}+\hat{R}z_{t}^{*}+\hat{R}^{\dagger}w_{t}^{*},\nonumber \\
 & -\hat{R}^{\dagger}\bar{O}_{1}-\hat{R}\bar{O}_{2}]|\psi(t,z^{*},w^{*})\rangle
\end{align}
where $\hat{R}=(\kappa_{1}\hat{q}_{1}+\kappa_{2}\hat{q}_{2})=\hat{R}^{\dagger}$
for this particular model and $z_{t}^{*}=-i\sum_{k}\sqrt{\bar{n}_{k}+1}g_{k}z_{k}^{*}e^{i\omega_{k}t}$,
$w_{t}^{*}=-i\sum_{k}\sqrt{\bar{n}_{k}}g_{k}w_{k}^{*}e^{-i\omega_{k}t}$
are two noise variables for those two effective baths after transformation.
The two operators $\hat{O}_{1}$ and $\hat{O}_{2}$ satisfy the relations
\begin{eqnarray}
\frac{\partial}{\partial t}\hat{O}_{1} & = & [-i\hat{H}_{S}+\hat{R}z_{t}^{*}+\hat{R}^{\dagger}w_{t}^{*}-\hat{R}^{\dagger}\bar{O}_{1}-\hat{R}\bar{O}_{2},\hat{O}_{1}]\nonumber \\
 &  & -\hat{R}^{\dagger}\frac{\delta}{\delta z_{s}^{*}}\bar{O}_{1}-\hat{R}\frac{\delta}{\delta z_{s}^{*}}\bar{O}_{2},
\end{eqnarray}
\begin{eqnarray}
\frac{\partial}{\partial t}\hat{O}_{2} & = & [-i\hat{H}_{S}+\hat{R}z_{t}^{*}+\hat{R}^{\dagger}w_{t}^{*}-\hat{R}^{\dagger}\bar{O}_{1}-\hat{R}\bar{O}_{2},\hat{O}_{2}]\nonumber \\
 &  & -\hat{R}^{\dagger}\frac{\delta}{\delta w_{s}^{*}}\bar{O}_{1}-\hat{R}\frac{\delta}{\delta w_{s}^{*}}\bar{O}_{2},
\end{eqnarray}
with the initial conditions $\hat{O}_{1}(t,s=t,z^{*},w^{*})=\hat{R}$
and $\hat{O}_{2}(t,s=t,z^{*},w^{*})=\hat{R}^{\dagger}$, where $\bar{O}_{i}=\int_{0}^{t}\alpha_{i}(t,s)\hat{O}_{i}(t,s,z^{*},w^{*})ds$
(i = 1, 2). It is easy to note that in our particular model $\hat{R}=(\kappa_{1}\hat{q}_{1}+\kappa_{2}\hat{q}_{2})=\hat{R}^{\dagger}$,
therefore two noises $z_{t}^{*}$ and $w_{t}^{*}$ as well as two
$\hat{O}$ operators $\hat{O}_{1}$ and $\hat{O}_{2}$ can be combined
as a single noise and a single $\hat{O}$ operator. Eventually, the
NMQSD equation still keeps the form of Eq.~(\ref{eq:QSD}), except
the correlation function is slightly modified. For details, see~\cite{PhysRevA.69.062107}.

\section{Chaos generation without optomenchanical couplings}
\label{sec:AppE}

\begin{figure} [t]
	\noindent\centering \includegraphics[width=1\columnwidth]{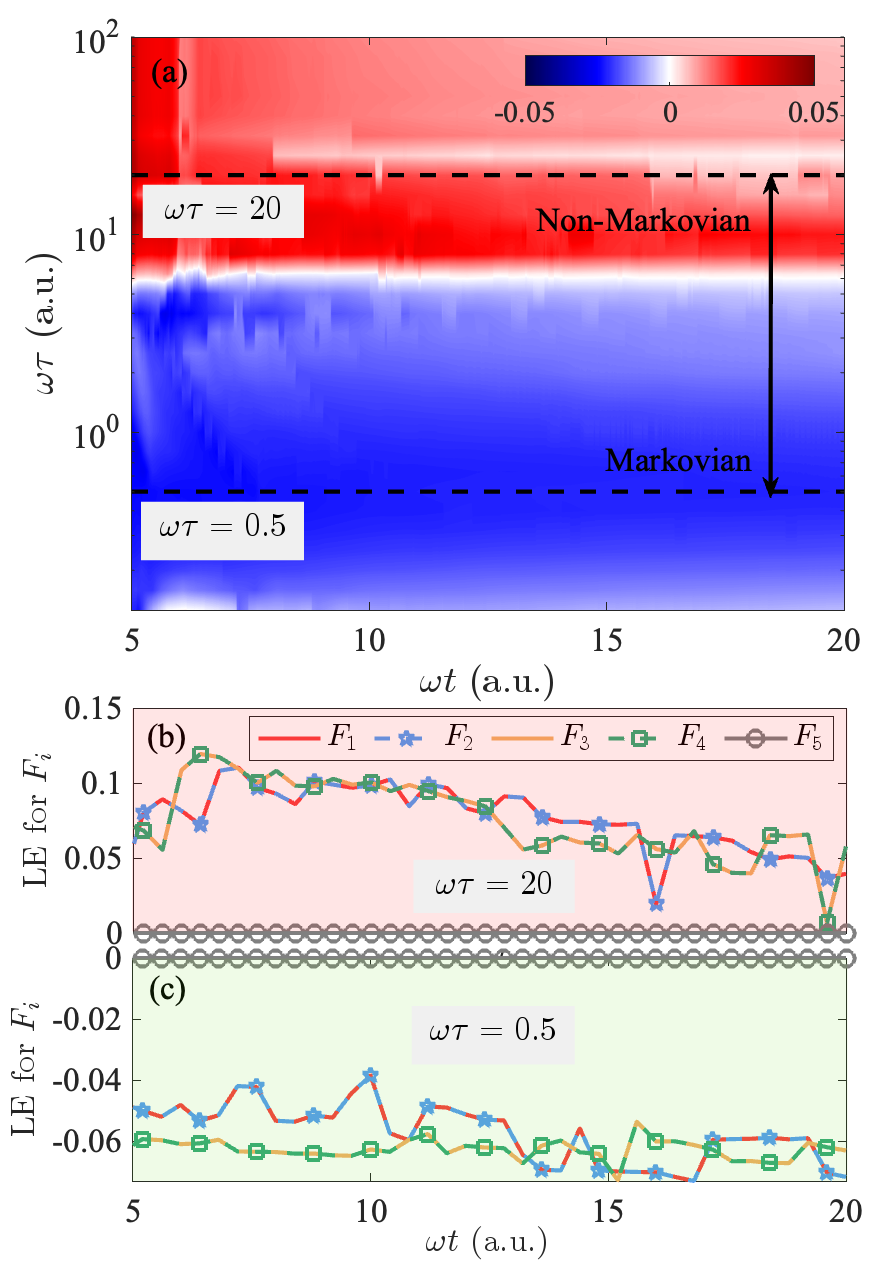}
	\caption{(a) The evolution of maximum LE for physical observable
		$\langle\hat{p}_{1}\rangle$ without optomechanical couplings ($G_{1}=G_{2}=0$). 
		(b) and (c) show the time evolution of maximum LE of the TDCs $F_{i}$ with given memory
		time $\tau$ marked in (a), respectively. The initial conditions are $\langle\hat{q}_{1}\rangle|_{t=0}=\langle\hat{q}_{2}\rangle|_{t=0}=0$
		and $\langle\hat{p}_{1}\rangle|_{t=0}=\langle\hat{p}_{2}\rangle|_{t=0}=1.1$
		and $\langle\hat{a}^{\dagger}\hat{a}\rangle|_{t=0}=2$. The parameters
		are $\omega_{1}=\omega_{2}=\omega=1$, $\Omega=0$,
		and $\kappa_{1}=\kappa_{2}=2.02$.}
	\label{fig6}
\end{figure}

Given the fact that the generation of chaos in optomechanical systems has been extensively studied \cite{Ma2023AdP,PhysRevA.81.013802,PhysRevLett.114.253601}, it is a well-established conclusion that optomechanical coupling can induce chaos. Consequently, upon obtaining the numerical results presented in Fig~\ref{fig22} and  Fig~\ref{fig3}, a natural question arises as to whether the chaotic phenomena depicted in these figures might be attributed to optomechanical couplings rather than being triggered by non-Markovian effects. Therefore, we present the numerical simulation for the case $G_1=G_2=0$ in Fig.~\ref{fig6}. The chaos generation without $G_1$ and $G_2$ excludes the possibility that chaos is originated from optomechanical couplings instead of non-Markovian feedback effects. Once more, the results in Fig.~\ref{fig6}(b)
and Fig.~\ref{fig6}(c) confirm our analysis in Sec.~\ref{III} that non-Markovian corrections provide non-linearity in TDCs thus cause chaotic dynamics eventually.

Here, we have to emphasize that our results demonstrate for the possibility of chaos induced purely by non-Markovian effects with strong evidence. But the examples shown in the paper, including $G=0$
and $G\neq0$, are only special cases with chaotic dynamics. Certainly, non-Markovian environments with feedback effects cannot always ensure chaotic dynamics. The condition for chaos induced by non-Markovian environment is still an open question.


%

\end{document}